\documentclass[12pt]{iopart}
\usepackage{graphicx}

\newcommand{\beq}{\begin{equation}}
\newcommand{\eeq}{\end{equation}}
\newcommand{\bea}{\begin{eqnarray}}
\newcommand{\eea}{\end{eqnarray}}
\renewcommand{\a}{\alpha}
\newcommand{\abs}[1]{\vert#1\vert}
\newcommand{\all}{{\rm all}}
\renewcommand{\b}{{\rm b}}
\renewcommand{\d}{{\rm d}}

\renewcommand{\e}{{\rm e}}

\newcommand{\eq}{{\rm eq}}

\renewcommand{\lim}{{\rm lim}}
\newcommand{\lin}{{\rm lin}}

\newcommand{\mean}[1]{\langle#1\rangle}
\newcommand{\s}{\sigma}

\newcommand{\tot}{{\rm tot}}

\newcommand{\D}{\Delta}

\eqnobysec

\begin{document}
\title[Statistics of bridge geometries]
{Cooperativity in sandpiles:\\ statistics of bridge geometries}
\author{Anita Mehta\dag, G C Barker\ddag, and J M Luck\S}

\address{\dag\ S N Bose National Centre for Basic Sciences, Block JD, Sector 3,
Salt Lake, Calcutta 700098, India}

\address{\ddag\ Institute of Food Research, Colney Lane, Norwich NR4 7UA, UK}

\address{\S\ Service de Physique Th\'eorique\footnote{URA 2306 of CNRS},
CEA Saclay, 91191~Gif-sur-Yvette cedex, France}

\begin{abstract}
Bridges form dynamically in granular media
as a result of spatiotemporal inhomogeneities.
We classify bridges as linear and complex,
and analyse their geometrical characteristics.
In particular,
we find that the length distribution of linear bridges is exponential.
We then turn to the analysis of the orientational distribution
of linear bridges and find that, in three dimensions,
they are {\it vertically diffusive but horizontally superdiffusive};
thus, when they exist, long linear bridges form `domes'.
Our results are in good accord with Monte Carlo simulations
of bridge structure;
we make predictions for quantities that are experimentally accessible,
and suggest that bridges are very closely related to force chains.
\end{abstract}
\eads{\mailto{anita@bose.res.in},
\mailto{barker@bbsrc.ac.uk},
\mailto{luck@spht.saclay.cea.fr}}
\maketitle

\section{Introduction}

The surface of a sandpile cloaks within it a vast array
of complex structures -- networks of grains whose stability
is interconnected, surrounded by pores and necks of void space.
Bridges -- arch-like structures, where mutual stabilisation
is a principal ingredient -- are prime among these, spanning
all manner of shapes and sizes through a granular medium.
They can be stable for arbitrarily long times, since the Brownian motion
that would dissolve them away in a liquid is absent in sandpiles
-- grains are simply too large for the ambient temperature
to have any effect.
As a result, they can affect
the ensuing dynamics of the sandpile; a major mechanism of compaction is
the gradual collapse of long-lived bridges in weakly
vibrated granular media, resulting
in the disappearance of the voids that were earlier enclosed~\cite{usbr}.
Bridges are also responsible for the `jamming'~\cite{refs} that
occurs, for example, as grains flow out of a hopper.

The difficulty of even identifying, leave alone analysing,
structures as complex as bridges
in a three-dimensional assembly should not be underestimated;
such an algorithm now exists, and
its use has resulted in the identification
and classification of a panoply of bridge configurations
generated via numerical simulations~\cite{I,II}.
However, even given the presence of such data,
it is a far from obvious task to cast it in theoretical terms.
In the following, we present a theory for the formation of bridges,
which, among other things, is able to explain at least
some of the salient features of our numerical data.

We first define a bridge.
Consider a stable packing of hard spheres under gravity, in three dimensions.
Each particle typically rests on three others, which stabilise it,
in the sense that downward motion is impeded.
{\it A bridge is a configuration of particles in which
the three-point stability conditions of two or more particles
are linked; that is, two or more particles are mutually stabilised}.
They are thus structures which cannot be formed by
the {\it sequential} placement of individual particles;
they are, however, frequently formed
by natural processes such as shaking and pouring,
where cooperative effects arise naturally.
In a typical packing, upto 70 percent of particles
are involved in bridge configurations.
The detailed history of a stabilisation process identifies a unique
set of bridges, since the stabilising triplet for each sphere is well defined.
While it is impossible to determine bridge distributions
uniquely from an array of coordinates representing particle positions,
we are able via our algorithm to obtain ones
that are the most likely to be the result of a given
stabilisation process~\cite{I,II}.

\begin{figure}[htb]
\begin{center}
\includegraphics[angle=0,width=.38\linewidth]{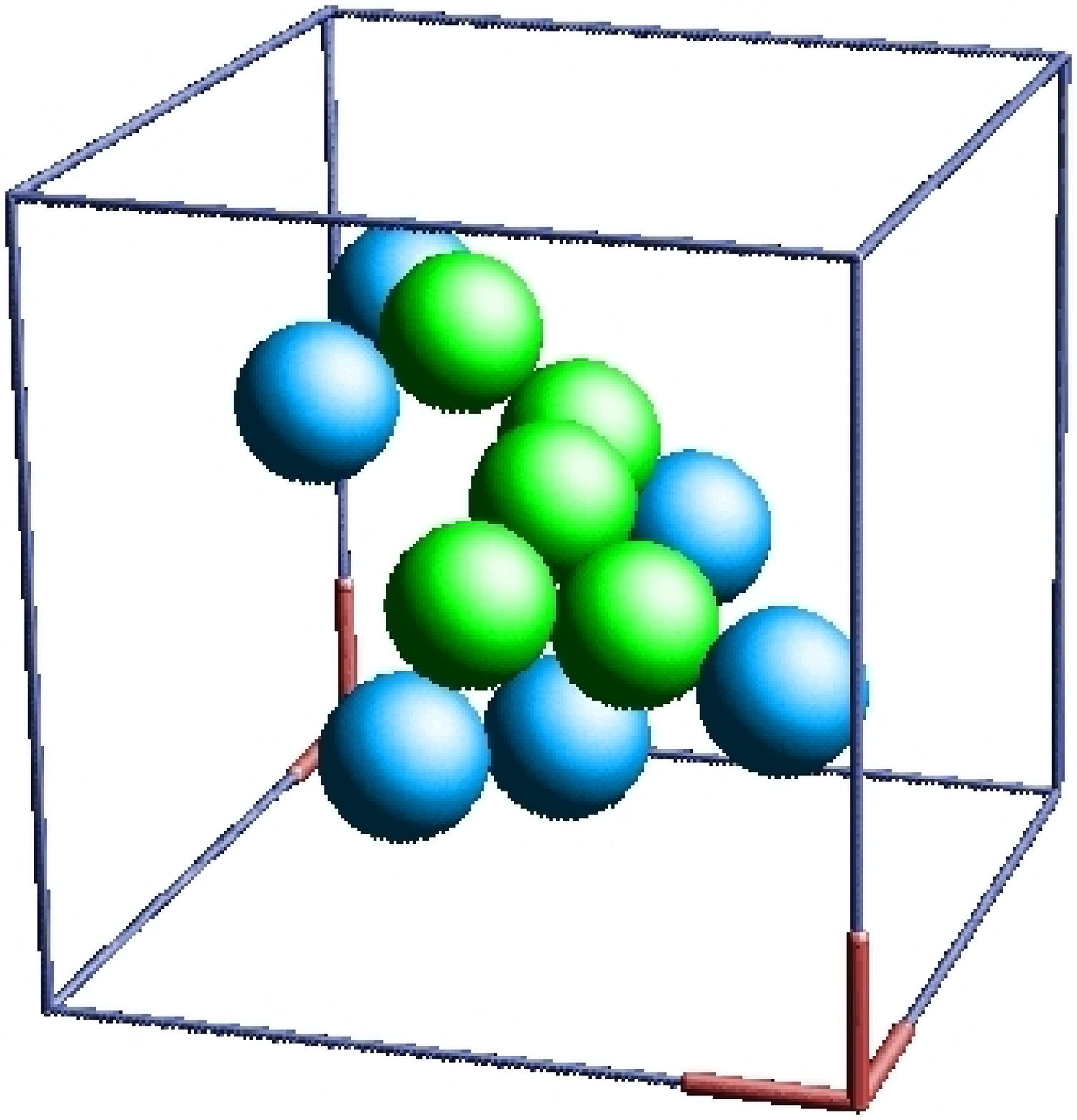}
\includegraphics[angle=0,width=.38\linewidth]{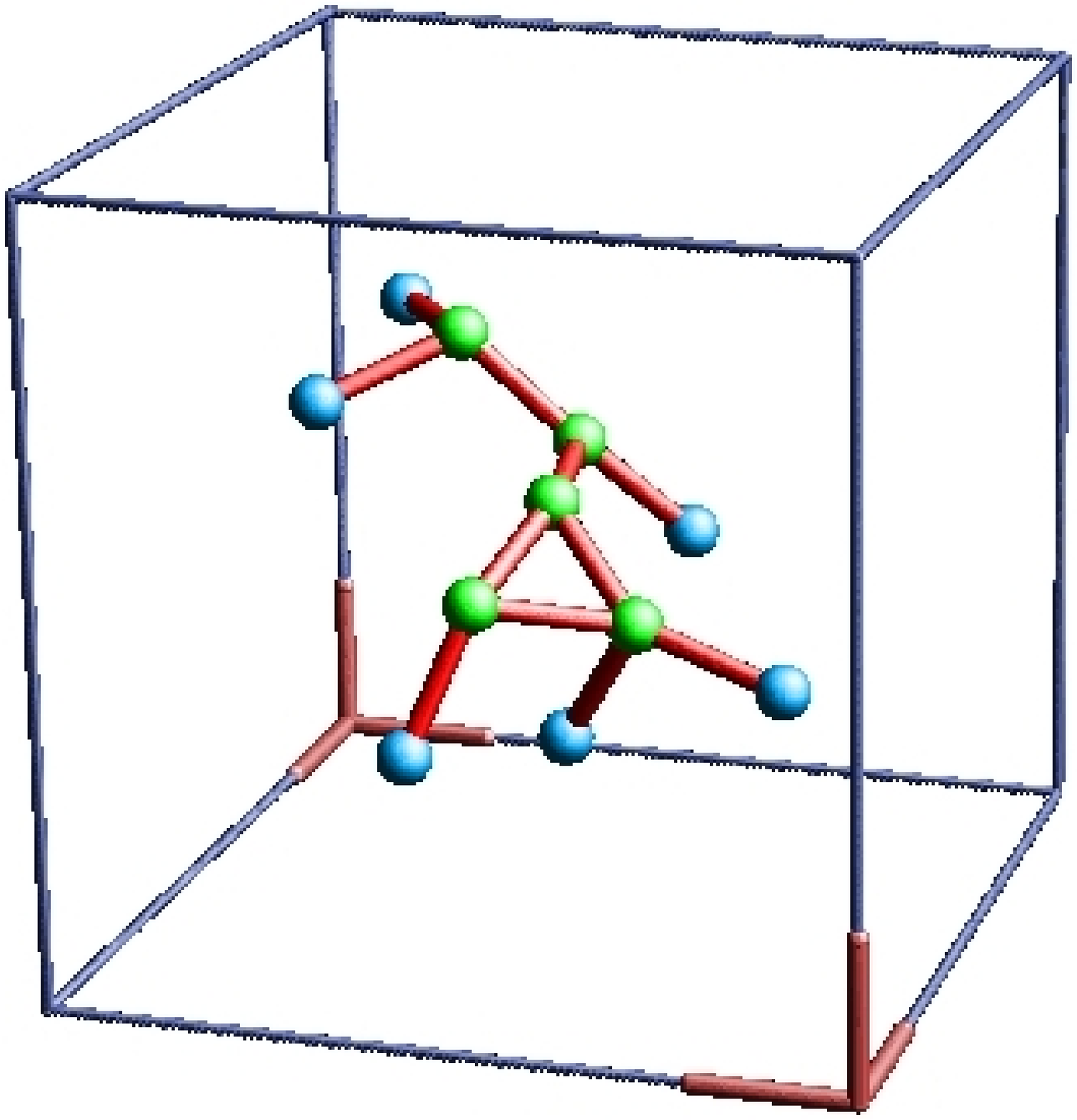}
\caption{\small
A five particle {\it complex bridge}, with six base particles (left),
and the corresponding contact network (right).
Thus $n=5$ and $n_\b=6<5+2$.}
\label{fig1}
\end{center}
\end{figure}

\begin{figure}[htb]
\begin{center}
\includegraphics[angle=0,width=.38\linewidth]{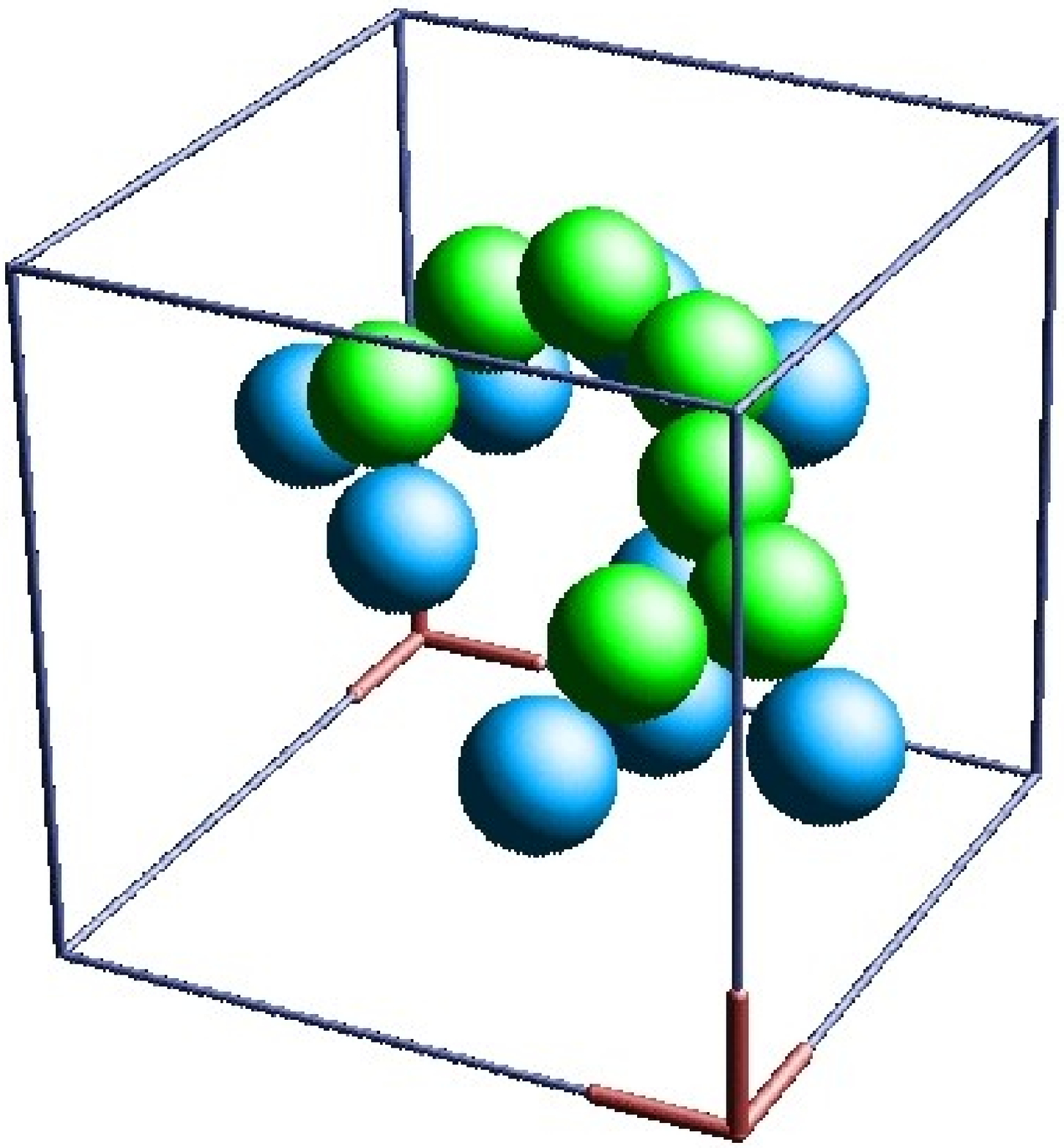}
\includegraphics[angle=0,width=.4\linewidth]{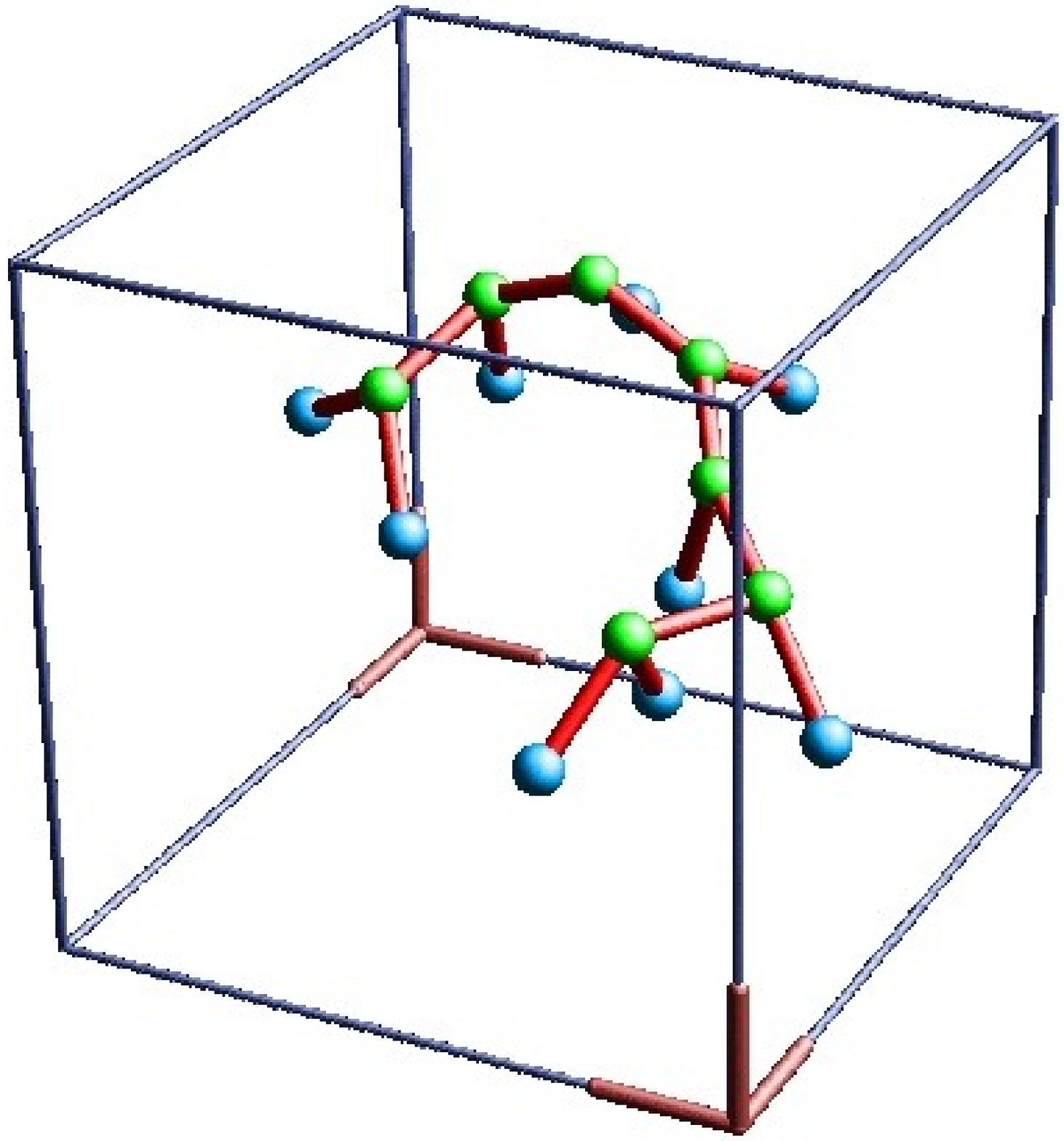}
\caption{\small
A seven particle {\it linear bridge} with nine base particles (left),
and the corresponding contact network (right).
Thus $n=7$ and $n_\b=9=7+2$.}
\label{fig2}
\end{center}
\end{figure}

We now distinguish between {\it linear} and {\it complex} bridges
via a comparison of Figures~\ref{fig1} and~\ref{fig2}.
Figure~\ref{fig1} illustrates a {\it complex} bridge,
i.e., a mutually stabilised cluster of five particles (shown in green),
where the stability is provided by six stable base particles (shown in blue).
Of course the whole is embedded in a stable network of grains within
the sandpile.
Also shown is the network of contacts for the particles in the bridge:
we see clearly that three of the particles each have two mutual stabilisations.
Figure~\ref{fig2} illustrates a seven particle linear bridge with nine
base particles.
This is an example of a {\it linear} bridge.
The contact network shows that this bridge
has a simpler topology than that in Figure~\ref{fig1}.
Here, all of the mutually stabilised particles are in sequence, as in a string.
A linear bridge made of $n$ particles therefore always rests
on $n_\b=n+2$ base particles,
whereas the number $n_\b$ of base particles of a complex bridge of size $n$
varies from one bridge to another, and always obeys $n_\b<n+2$,
because of the presence of loops in the contact network of complex bridges.

An important point to note is that bridges can only be formed sustainably
in the presence of friction; the mutual stabilisations needed would
be unstable otherwise! Although we use Monte Carlo
simulations (described below) which do not contain friction
explicitly, the configurations we generate correspond to those
in nature that include frictional effects.
In particular, they generate coordination numbers in a range that is consistent
with the presence of friction~\cite{usbr,sam,silbert}.
Additionally, we have analysed the configurations of molecular dynamics
simulations in the limit of high friction\footnote{We thank
Leo Silbert for letting us use configurations generated from his
molecular dynamics simulations, where we have found and analysed
the distribution of bridges.};
these generate distributions of bridges very similar to our own.
Our simulations suggest strong analogies
between bridge structure and force distributions in granular media,
which have been measured both in experiments~\cite{mueth}
and in simulations~\cite{sid}.
This indicates that bridges might be com\-ple\-men\-tary
(and experimentally easier) probes of inhomogeneities in these systems.

\section{Simulation details}

We have examined bridge structures in hard assemblies that are
generated by an established, non-sequential,
restructuring algorithm~\cite{usbr,I,II},
whose main modelling ingredients involve {\it stochastic} grain displacements
and {\it collective} relaxation from them.

This algorithm restructures a stable
hard sphere deposit in three distinct stages.

\begin{itemize}

\item[(1)]
The assembly is dilated in a vertical direction (with free volume
being introduced homogeneously throughout the system), and each
particle is given a random horizontal displacement; this models
the dilation phase of a vibrated granular medium.

\item[(2)]
The packing is compressed in a uniaxial external field representing gravity,
using a low-temperature Monte Carlo process.

\item[(3)]
The spheres are stabilised using a steepest descent `drop and roll' dynamics to
find a local minimum of the potential energy.

\end{itemize}
Steps (2) and (3) model the quench phase of the vibration,
where particles relax to locally stable positions in the presence of gravity.
Crucially, during the third phase, the spheres
are able to roll in contact with others; {\it mutual stabilisations} are thus
allowed to arise, mimicking collective effects.
The final configuration has
a well-defined network of contacts and each sphere is supported,
in its locally stable position, through point contacts, by a set
of three other spheres uniquely defined.
In practice, the final
configuration may include a few non-stabilised particles.

The simulation method recalled above builds a sequence of static packings.
Each new packing is built from its predecessor by a random process and
the sequence achieves a steady state.
In the steady state thus obtained, structural descriptors such as the
mean packing fraction and the mean coordination number
fluctuate about well-defined mean values.
The steady-state mean volume fraction $\Phi$
is found to lie typically in the range $\Phi\sim 0.55\hbox{ -- } 0.61$.
This volume fraction depends on all the parameters of the simulation,
and can therefore be adjusted to any desired value in the above range.
The mean coordination number is always $Z\approx 4.6\pm 0.1$.
Since for frictional packings the minimal coordination number is
$Z=d+1$~\cite{sam}, this confirms that our 3D configurations
correspond to those generated in the presence of friction.
In fact, the mean coordination number
of molecular dynamics configurations of frictional sphere packings
is slightly above 4.5, in the limit of a large
friction coefficient~\cite{silbert}.
We recall that, in the absence of friction, the coordination number
is that of an isostatic system, $Z=2d$~\cite{donev},
hence $Z=6$ in the present three-dimensional situation.

To be more specific, simulations were performed in a rectangular cell
with lateral periodic boundaries, and a hard disordered base.
The simulations, performed
serially on a desktop workstation, have a very time-consuming
Monte Carlo compression phase with each simulation taking
several days' worth of CPU time.
This is necessary in order to have simulation results which are
reproducible, without appreciable dependence on
Monte Carlo parameters or system size.
Each of our configurations includes about $N_\tot=2200$ particles.
We have examined ap\-pro\-xi\-ma\-tely $100$ configurations from the
steady states of the reorganisation process
for the following two values of the packing fraction:
$\Phi=0.56$ and $\Phi=0.58$.
Segregation is avoided by choosing monodisperse particles; a rough
base prevents ordering.
A large number of restructuring cycles is needed to
reach the steady state for a given shaking amplitude;
about 100 stable configurations (picked every 100
cycles in order to avoid correlation effects) are saved for future analysis.
{}From these configurations, and following
specific prescriptions, our algorithm identifies bridges
as clusters of mutually stabilised particles~\cite{I,II}.

\begin{figure}[htb]
\begin{center}
\includegraphics[angle=0,width=.6\linewidth]{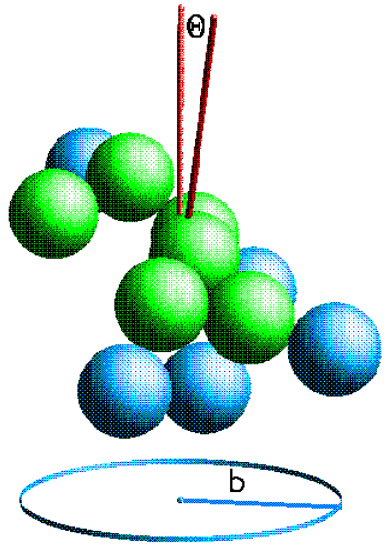}
\caption{\small
Definition of the angle $\Theta$ and the base extension $b$ of a bridge.
The main axis makes an angle $\Theta$ with the $z$-axis;
the base extension $b$ is the radius of gyration of the base particles
about the $z$-axis.}
\label{figbig}
\end{center}
\end{figure}

Figure~\ref{figbig} illustrates two characteristic descriptors
of bridges used in this work.
Along the lines of~\cite{II}, the {\it main axis} of a bridge
is defined using a triangulation of its base particles.
Triangles are constructed by choosing all possible connected triplets
of base particles: the vector sum of their normals
is defined to be the direction of the main axis of the bridge.
The orientation angle $\Theta$ is defined as the angle
between the main axis and the $z$-axis.
The {\it base extension} $b$ is defined as the radius of gyration of
the base particles about the $z$-axis (which is thus distinct
from the radius of gyration about the main axis of the bridge).

\section{Geometrical characteristics.~Size and diameter distribution}

In the following, we present statistics
for both linear and complex bridges; however it is the former that we analyse
in greater detail, both because they are conceptually simpler and
because they are in fact more numerous in our simulations.
The latter clearly show that as a linear bridge
gets longer, it is increasingly likely to develop
branches, whereupon it becomes complex~\cite{I,II}.

The formation of either type of bridge in shaken granular
media is, of course, a dynamical process; we, however,
adopt an ergodic viewpoint~\cite{edwards} here, inspired by the
ensemble statistics of bridges generated by our simulations.
As in polymer theory~\cite{doi}, we differentiate between
linear and complex geometries; in our phenomenology, we regard a linear bridge
as a random chain which
grows in arc length $s$ by the sequential addition of links to its end,
while complex bridges, being branched structures, cannot grow sequentially.
This replacement of what is in reality a collective phenomenon by
a random walk is somewhat analogous to the `tube model'
of linear polymers~\cite{doi}: both are simple but efficient effective
pictures of very complex problems.

In this spirit, we first address the question of the length distribution
of linear bridges.
We define the length distribution $f_n$ as the probability that a linear bridge
consists of exactly $n$ spheres.
We make the simplest and the most natural assumption
that a bridge of size $n$ remains linear with some probability $p<1$
if an $(n+1)^{\rm th}$ sphere is added to it, in the above sense.
This simple effective picture therefore leads to the exponential distribution
\beq
f_n=(1-p)p^n.
\label{fexp}
\eeq
This exponential law is a common feature of all models
for random `strings' made of~$n$ units, irrespective of the way they are formed.
An illuminating example is provided by percolation clusters
in one dimension, which are the prototype of random linear objects~\cite{stah}.

The above argument leading to an exponential distribution
can alternatively be reformulated as follows, by means of a continuum approach.
This will give us the opportunity
to introduce a formalism that will be extensively used later.
A linear bridge is now viewed as a continuous random curve or `string',
parametrised by the arc length~$s$ from one of its endpoints.
Within the above picture of a sequential process,
we assume that a linear bridge
disappears at a constant rate $\a$ per unit length,
either by changing from linear to complex or by collapsing.
The probability $S(s)$ that a given bridge survives
at least up to length~$s$ thus obeys the rate equation
$\dot S=-\a S$, and therefore falls off exponentially,
according to $S(s)=\exp(-\a s)$.
The probability distribution of the length $s$ of linear bridges
therefore reads $f(s)=-\dot S(s)=\a\,\exp(-\a s)$.
This is the continuum analogue of~(\ref{fexp}).

\begin{figure}[htb]
\begin{center}
\includegraphics[angle=90,width=.58\linewidth]{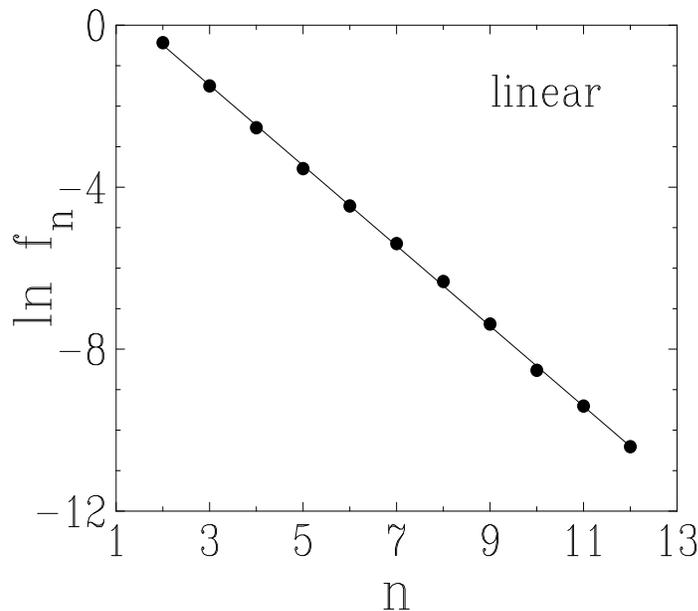}
\caption{\small
Logarithmic plot of the length distribution $f_n$ of linear bridges,
against $n$, for $\Phi=0.58$.
Full line: least-square fit yielding $p\approx0.37$, i.e., $\a\approx0.99$.}
\label{figflin}
\end{center}
\end{figure}

Figure~\ref{figflin} shows a logarithmic plot of numerical data
for the length distribution $f_n$ of linear bridges.
The data exhibit an exponential fall off of the form~(\ref{fexp}),
with $p\approx0.37$, i.e., $\a\approx0.99$\footnote{Despite its proximity
to unity, this value of $\a$ has no special significance, as can
be seen from the corresponding value of $p$.}.
This exponential decay of the distribution of linear bridges is
very accurately observed until $n\approx12$.

\begin{figure}[htb]
\begin{center}
\includegraphics[angle=90,width=.6\linewidth]{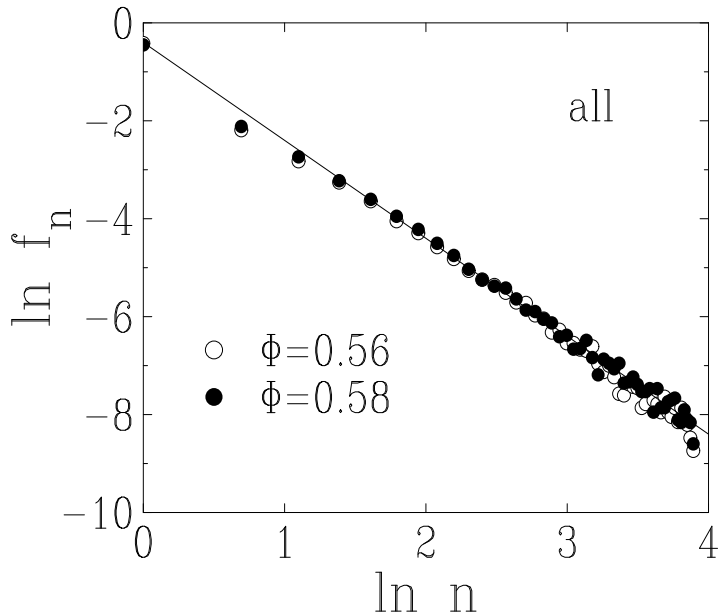}
\caption{\small
Log-log plot of the size distribution $f_n$ of all bridges,
against $n$, for two volume fractions.
This
distribution is dominated by complex bridges, since
these predominate after $n\sim 8$.
Full line: least-square fit yielding $\tau\approx2.00$.}
\label{figfall}
\end{center}
\end{figure}

Complex bridges begin to be dominant around $n\approx8$.
For a complex bridge, the number $n$ is now referred to
as the `size' of the bridge.
As complex bridges are branched objects,
it is natural to expect that their size distribution resembles
that of generic randomly branched objects,
such as e.g.~lattice animals or critical percolation clusters~\cite{stah}.
These prototypical examples have a power-law fall-off
in their size distribution:
\beq
f_n\sim n^{-\tau}.
\label{fpower}
\eeq
The numerical data shown in Figure~\ref{figfall}
clearly show that the size distribution of bridges obeys
a power-law behaviour of the form~(\ref{fpower}),
and thus behave as generic randomly branched objects.
The measured value of the exponent $\tau$
seems to coincide with the limiting value $\tau=2$,
at and below which the mean size $\mean{n}=\sum n\,f_n$ is divergent.

We now turn to the diameter of linear and complex bridges,
which is one of the most important of their geometrical characteristics.
The typical diameter $R_n$ of a bridge of size $n$
is such that $R_n^2$ is the mean squared end-to-end distance
over all the bridges of size $n$.
We examine this quantity
separately for linear bridges, as well as the ensemble of all bridges.
In both cases we expect and find a power-law behaviour
\beq
R_n\sim n^\nu.
\label{nudef}
\eeq

\begin{figure}[htb]
\begin{center}
\includegraphics[angle=90,width=.54\linewidth]{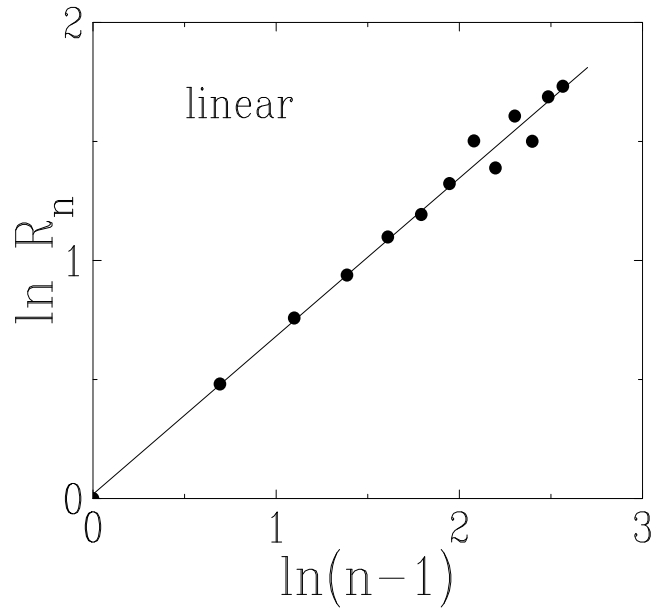}
\caption{\small
Log-log plot of the typical diameter $R_n$ of linear bridges,
against $n-1$, the number of mutually stabilising bonds, for $\Phi=0.58$.
Full line: least-square fit yielding $\nu_\lin\approx0.66$,
i.e., $D_\lin\approx1.51$.}
\label{figrlin}
\end{center}
\end{figure}

For linear bridges, $D_\lin=1/\nu_\lin$ can be interpreted as their
fractal dimension.
The observed value (see Figure~\ref{figrlin}) $\nu_\lin\approx0.66$,
i.e., $D_\lin\approx1.51$,
lies between those for a self-avoiding walk in two ($\nu_2=3/4$)
and three ($\nu_3\approx0.59$) dimensions~\cite{saw}.
This seems entirely reasonable, since linear bridges
are likely to behave as typical self-avoiding curves.
Furthermore, in three dimensions, their effective dimensionality
is expected to lie between 2 and 3: they may
start off confined to a plane, and then collapse onto each other, due to the
effects of vibration.

\begin{figure}[htb]
\begin{center}
\includegraphics[angle=90,width=.55\linewidth]{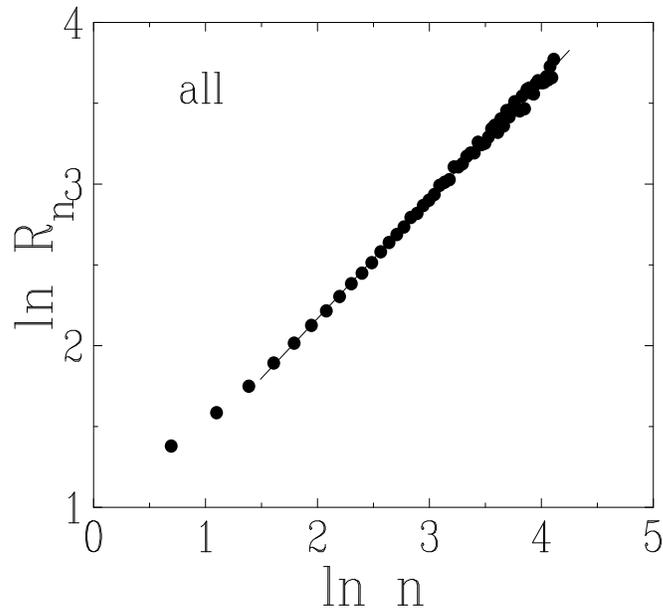}
\caption{\small
Log-log plot of the typical diameter $R_n$ of all bridges, against size $n$,
for $\Phi=0.58$.
These statistics mainly derive from complex
bridges, which dominate beyond about $n\approx8$.
Full line: least-square fit to points with $n>5$,
yielding $\nu_\all\approx0.74$.}
\label{figrall}
\end{center}
\end{figure}

For complex bridges, we again find a power law of the form~(\ref{nudef}).
The geometrical exponent $\nu_\all\approx0.74$ (see Figure~\ref{figrall})
is now naturally compared with the value
$\nu\approx0.875$ of 3D percolation~\cite{stah}.
(Although the figure shows statistics for `all' bridges,
these mostly reflect the behaviour of complex bridges,
which dominate beyond about $n\approx8$.)

\begin{figure}[htb]
\begin{center}
\vskip 10pt
\includegraphics[angle=0,width=.6\linewidth]{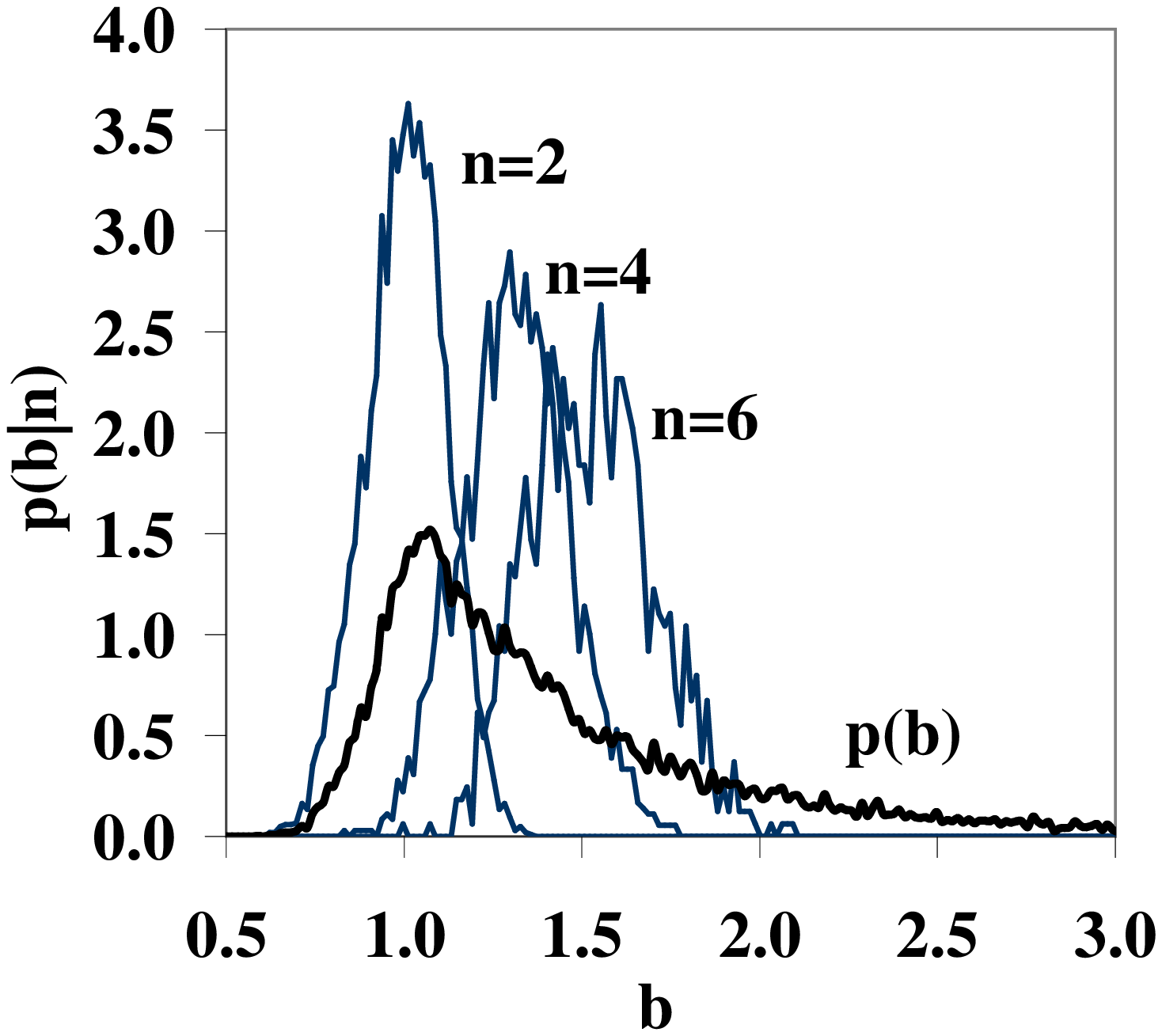}
\vskip 10pt
\includegraphics[angle=0,width=.6\linewidth]{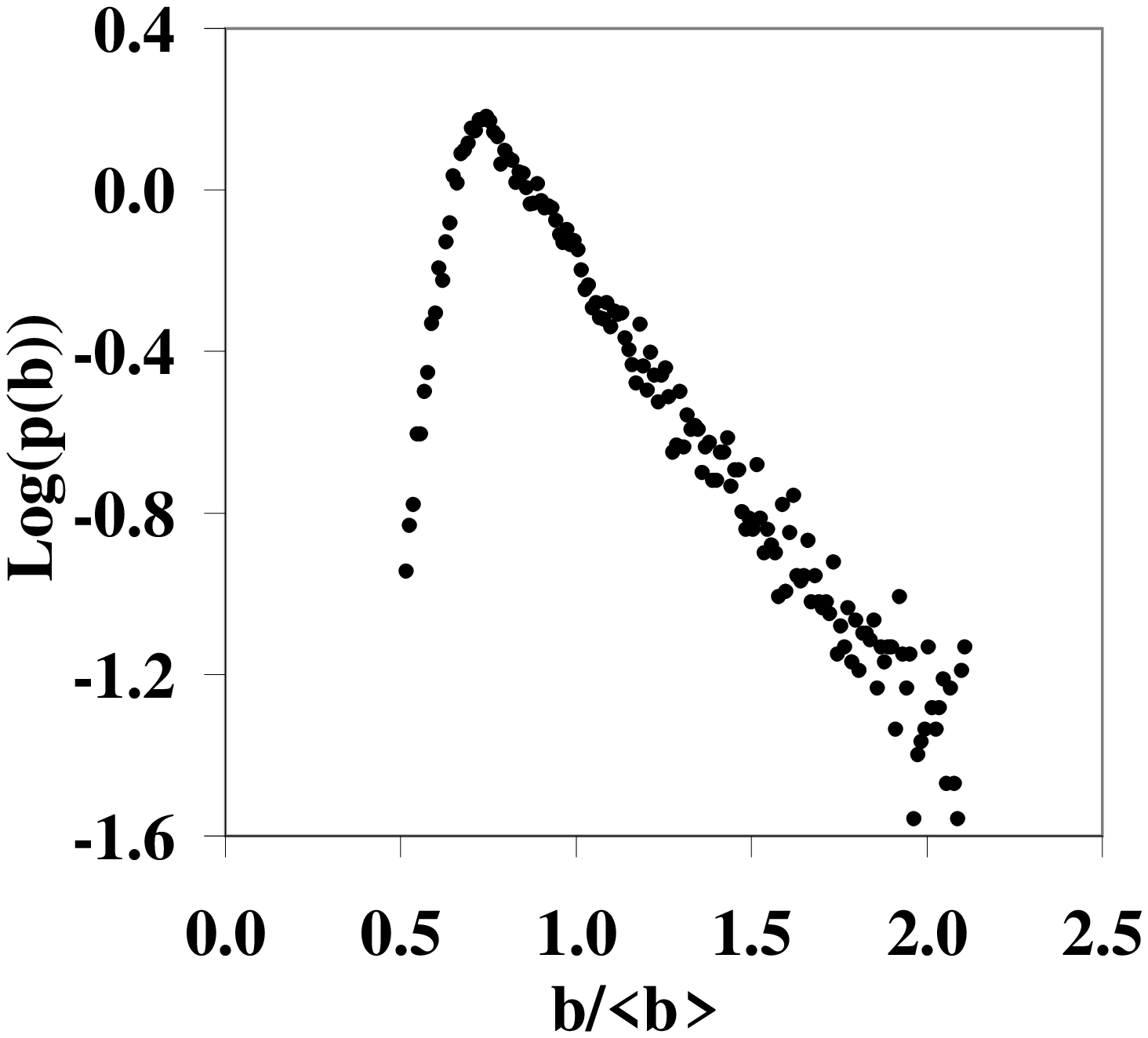}
\caption{\small
Distribution of base extensions of bridges, for $\Phi=0.58$.
In the top panel, the distributions
conditional on the bridge size $n$, for $n=2,4,6$ are marked in blue,
while the cumulative distribution is drawn in black.
The bottom panel shows the logarithm of the normalised probability distribution,
as a function of the normalised variable $b/\mean{b}$.}
\label{fig8}
\end{center}
\end{figure}

Lastly, we plot in Figure~\ref{fig8} the probability distribution function of
the base extension~$b$, for linear bridges; this
is clearly a measure of the spanning, and hence the jamming,
potential of a bridge (see Figure~\ref{figbig}).
We plot also in Figure~\ref{fig8} distribution functions
that are conditional on the bridge size $n$.
This was done to compare our results with those of~\cite{pak},
bearing in mind of course that our results
refer to sphere packings in three dimensions whereas
theirs refer to disk packings in two.
In our case, the conditional distributions are sharply peaked,
and only extend over a finite range of~$b$;
in theirs the peaks are smaller and the distributions broader.
Our results indicate that at least in three dimensions, and upto
the lengths of linear bridges probed, bridges of a given length
$n$ have a fairly characteristic horizontal extension; we can predict
fairly reliably the dimension of the orifice that they might
be expected to jam.
The results of~\cite{pak} indicate,
on the contrary, that there is a wide range of orifices
which would be jammed by disk packings of a given length, in two dimensions.
Also, for our three-dimensional bridges,
the cumulative distribution has a long tail at
large extensions, reflecting chiefly the existence of larger bridges
than in the data of~\cite{pak}.
(We note that this long tail
is characteristic of three-dimensional experiments
on force chains~\cite{mueth,ohern} which we will discuss
at greater length below.)
All of this seems entirely reasonable
given the more complex topology of three-dimensional space,
as well as the greater diversity of possible sphere packings within it.

In the lower part of Figure~\ref{fig8} we have plotted the logarithm
of the normalised probability distribution of base extensions,
against the normalised variable
$b/\mean{b}$, where~$\mean{b}$ is the mean extension of bridge bases.
This figure emphasises the exponential tail of the distribution function,
and also shows that bridges with small base extensions are unfavoured.
It is interesting to note that the form of the normalised distribution
resembles that found in force distributions of emulsions~\cite{jasna},
and that there are also similarities with normal force distributions
in MD simulations of particle packings (see~e.g.~\cite{sid,ohern}).
In our simulations, the sharp drop at the origin
with a very tangible peak strongly resembles that obtained
in the latter case, in the limit of strong
deformations~\cite{sid}.
Realising that the measured
forces~\cite{sid,jasna} propagate through chains of particles,
we use this similarity to suggest that our bridges are
really just long-lived force chains, which have survived
despite strong deformations.
We suggest also that with the current availability of 3D
visualisation techniques such as NMR~\cite{fukushima},
bridge configurations might be an easily measurable,
and effective, tool to probe inhomogeneous force networks in shaken sand.

\section{Orientational distribution of linear bridges.
Theory and simulation}

Linear bridges predominate at small $n$, which is
after all the regime of interest for finite-sized simulations and experiments.
The theory we present below focuses on such bridges; the distribution
of their orientation is predicted, and then compared
with our simulation results.

As before, a linear bridge is modelled as a continuous curve,
parametrised by the arc length~$s$ from one of its endpoints.
We propose the following effective description of the local
and global orientation of this curve.
For the sake of simplicity,
we choose to focus on the most important degree of freedom,
which is the tilt with respect to the horizontal; the azimuthal degree
of freedom is therefore neglected.
Accordingly, we define the local or `link' angle $\theta(s)$
between the direction of the tangent to the bridge at point~$s$
and the horizontal,
and the mean angle made by the bridge from its origin up to point~$s$,
\beq
\Theta(s)=\frac{1}{s}\int_0^s\theta(u)\,\d u.
\label{Thdef}
\eeq
The local angle $\theta(s)$ so defined may be either
positive and negative; it can even change sign
along the random curve which represents a linear bridge.
Of course, the orientation angle $\Theta$ measured in our numerical simulations
pertains to the three-dimensional realm of real sphere packings.
It is there positive by construction,
being defined at the end of Section~2 as the angle between the main
bridge axis and the $z$-axis.

Our simulations show that
the angle $\Theta(s)$ typically becomes smaller and smaller
as the length $s$ of the bridge increases.
Small linear bridges are almost never flat~\cite{I,II}; as they get
longer, assuming that they still stay linear, they get `weighed down', arching
over as at the mouth of a hopper~\cite{br}.
Thus, in addition to our earlier claim that long
linear bridges are rare, we claim further here that
(if and) when they exist, they typically have flat bases, becoming `domes'.

We use these insights to write down equations
to investigate the angular distribution of linear bridges.
These couple the evolution of the local angle $\theta(s)$
with local density fluctuations $\phi(s)$ at point~$s$:
\bea
&&\dot\theta=-a\theta-b\phi^2+\D_1\eta_1(s),
\label{cou1}\\
&&\dot\phi=-c\phi+\D_2\eta_2(s).
\label{cou2}
\eea
The effects of vibration on each of $\theta$ and $\phi$ are represented by
two independent white noises $\eta_1(s)$, $\eta_2(s)$, such that
\beq
\mean{\eta_i(s)\eta_j(s')}=2\,\delta_{ij}\,\delta(s-s'),
\label{white}
\eeq
whereas the parameters $a$,~...,~$\D_2$ are assumed to be constant.

The phenomenology behind the above equations is the following:
the evolution of~$\theta(s)$ is caused by the addition to the bridge
of single particles to the bridge,
along the lines of our effective sequential picture.
The motion of particles within their cages gives rise to the fluctuations
of local density at a point $s$; thus $\phi$ can be regarded
as an effective {\it collective} coordinate,
with~$\theta$ seen as an {\it independent-particle} coordinate~\cite{mln}.

The individual dynamics are simple: the
first terms on the right-hand side of~(\ref{cou1}),~(\ref{cou2})
say that neither $\theta$ nor $\phi$ is allowed to be arbitrarily large.
Their coupling via the second
term in~(\ref{cou1}) arises as follows: large link
angles $\theta$ will be increasingly unstable in the
presence of density fluctuations $\phi^2$ of large magnitude,
which would to a first approximation `weigh the bridge down', i.e.,
decrease the angle $\theta$ locally.
Density fluctuations (otherwise known as `dilatancy')
also play a key role in the dynamics of
the angle of repose of a sandpile, which are described in a concurrent
paper~\cite{angle}.
There, sandpile collapse is described
in terms of an activated process, where an effective temperature
competes against configurational barriers generated by dilatancy.

Reasoning as above, we therefore anticipate that for low-intensity vibrations
and stable bridges, both density fluctuations $\phi(s)$
and link angles $\theta(s)$ will be small.
Accordingly, we linearise~(\ref{cou1}),
obtaining thus an Ornstein-Uhlenbeck equation~\cite{ou,vank}
\beq
\dot\theta=-a\theta+\D_1\eta_1(s).
\label{ou}
\eeq

Let us make the additional assumption that the initial angle $\theta_0$,
i.e., that observed for very small bridges,
is itself Gaussian with variance $\s_0^2=\mean{\theta_0^2}$.
The angle $\theta(s)$ is then a Gaussian process with zero mean
for any value of the length $s$.
Its correlation function can be easily evaluated to be
\beq
\mean{\theta(s)\theta(s')}
=\s_\eq^2\,\e^{-a\abs{s-s'}}+(\s_0^2-\s_\eq^2)\e^{-a(s+s')}.
\label{thcor}
\eeq
The characteristic length for the decay of orientation
correlations is $\xi=1/a$, and the variance of the link angle reads
\beq
\mean{\theta^2}(s)=\s_\eq^2+(\s_0^2-\s_\eq^2)\e^{-2as},
\eeq
where
\beq
\s_\eq^2=\frac{\D_1^2}{a}
\eeq
is the `equilibrium' value of the variance of the link angle.
Thus
as the chain gets longer, the variance of the link angle relaxes
from $\s_0^2$ (that for the initial link) to $\s_\eq^2$, in the limit
of an infinite bridge.

We predict from the above that the mean angle $\Theta(s)$
will have a Gaussian distribution, for any fixed length $s$.
By inserting~(\ref{thcor}) into~(\ref{Thdef}), we derive its variance:
\beq
\mean{\Theta^2}(s)=2\s_\eq^2\,\frac{as-1+\e^{-as}}{a^2s^2}
+(\s_0^2-\s_\eq^2)\frac{(1-\e^{-as})^2}{a^2s^2}.
\label{Thvar}
\eeq
The asymptotic result
\beq
\mean{\Theta^2}(s)\approx\frac{2\s_\eq^2}{as}\approx\frac{2\D_1^2}{a^2s}
\label{thas}
\eeq
confirms our earlier statement (see below~(\ref{Thdef}))
that a typical long bridge has a base that is almost
flat.
It can be viewed as consisting of a large number $as=s/\xi\gg1$ of
independent `blobs'\footnote{This is qualitatively reminiscent
of a similar problem in polymers,
where the de Gennes picture of `blob' dynamics in dilute solutions
gives way to rigid rod-like behaviour -- see e.g.~\cite{doi}.},
each of length $\xi$.

The result~(\ref{thas}) has another interpretation.
As $\Theta(s)$ is small with high probability for a very long bridge,
its extension in the vertical direction reads approximately
\beq
Z=z(s)-z(0)\approx s\,\Theta(s),
\label{zed}
\eeq
so that $\mean{Z^2}\approx s^2\mean{\Theta^2}(s)\approx2(\D_1/a)^2\,s$.
Going back to the discrete formalism, we have therefore
\beq
Z_n\sim n^{1/2}.
\eeq
The vertical extension of a linear bridge is thus found to grow with the usual
random-walk exponent $1/2$, whereas its horizontal extension
exhibits the non-trivial exponent $\nu_\lin\approx0.66$ of
Figure~\ref{figrlin}.
Thus, {\it long linear bridges are domelike; they are
vertically diffusive but horizontally superdiffusive}.
We recall
that the two-dimensional arches
found in~\cite{pak} were diffusive in a vertical
direction; our present results show that in three
dimensions, the vertical diffusivities of bridge structures
remain, and are enriched by a horizontal superdiffusivity.
Evidently, jamming in a three-dimensional hopper would be caused
by the planar projection of such a {\it dome}.

\begin{figure}[htb]
\begin{center}
\includegraphics[angle=90,width=.57\linewidth]{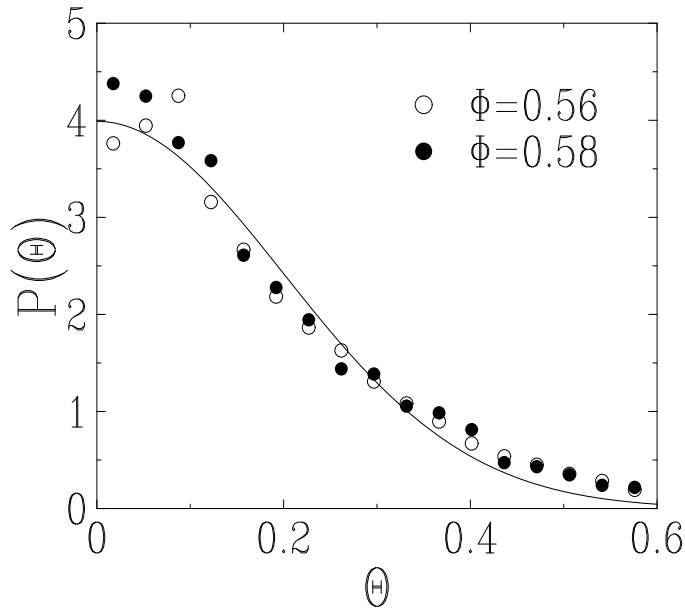}
\caption{\small
Plot of the normalised distribution of the mean angle $\Theta$
(in radians) of linear bridges of size $n=4$, for both volume fractions.
The $\sin\Theta$ Jacobian has been duly divided out,
explaining thus the larger statistical errors at small angles.
Full lines: common fit to (half) a Gaussian law.}
\label{figgauss}
\end{center}
\end{figure}

\begin{figure}[htb]
\begin{center}
\includegraphics[angle=90,width=.62\linewidth]{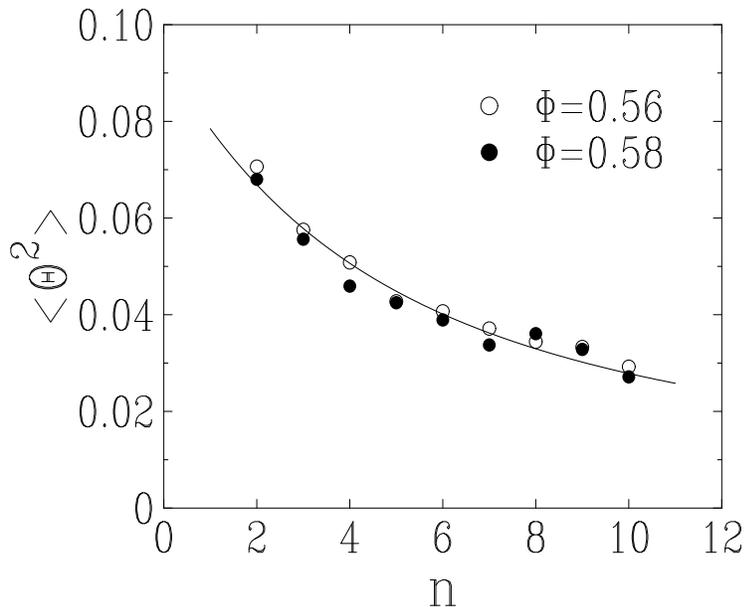}
\caption{\small
Plot of the variance of the mean angle of a linear bridge,
against size~$n$, for both volume fractions.
Full line: common fit to the first (stationary) term of~(\ref{Thvar}),
yielding $\s_\eq^2=0.093$ and $a=0.55$.
The `transient' effects of the second term of~(\ref{Thvar}) are invisible
with the present accuracy.}
\label{figvar}
\end{center}
\end{figure}

We now compare these predictions with data from our simulations.
The numerical data shown in Figure~\ref{figgauss}
confirm that the mean angle has, to a good approximation,
a Gaussian distribution in this typical situation of bridges of size $n=4$.
We recall that the angle $\Theta$ measured in simulations
pertains to the three-dimensional world.
Hence only the positive half of the Gaussian is used for comparison
with results of simulations, and the $\sin\Theta$
three-dimensional Jacobian is divided out.
It has also been checked to high accuracy
(less than 2 percent for a bin size $\delta\varphi=5$ degrees)
that the azimuthal angle $\varphi$
of the main bridge axis has the expected flat distribution.
Figure~\ref{figvar} shows the measured size dependence
of the variance $\mean{\Theta^2}(s)$, for both volume fractions.
The numerical data are found to agree well with a common fit
to the first (stationary) term of~(\ref{Thvar}),
with a common value of the parameters $\s_\eq^2=0.093$ and $a=0.55$.
The `transient' effects of the second term of~(\ref{Thvar}) are invisible
with the present accuracy (see below).
This observed agreement therefore provides a first confirmation
of the validity of our theory, quite remarkable given
that it is entirely independent of the simulations.
We conclude that in spite of its simplicity,
our theory captures the main features of linear bridges.

We end this section with the following remarks.
First, more subtle effects,
including the effects of transients via the second term of~(\ref{Thvar}),
and the dependence of the parameters $\s_\eq^2$ and~$a$
on the packing fraction $\Phi$,
are deserving of further investigation.
As a preface to our second remark,
we recall that in our phenomenological picture
linear bridge formation proceeds sequentially,
while complex bridges with their branched structures
form collectively, in space.\footnote{We reiterate once again
that this effective picture replaces the complexity
of the full dynamical picture, where {\it all} bridges form cooperatively.}
We might expect that with increasing density~$\Phi$, branched structures
would become more and more common; linear bridge formation, with its
`sequential' progressive attachment of independent blobs would then
become more and more rare.
Our theory should therefore cease to hold
at a limit packing fraction~$\Phi_\lim$,
which is reminiscent of the {\it single-particle
relaxation threshold density}~\cite{johannes},
at which the dynamics of granular compaction crosses over from being
single-particle (sequential) to~collective.
Finally, while we have
allowed for the local variability of density fluctuations
$\phi$ via~(\ref{cou2}), we have so far not explicitly coupled
these to bridge orientations.
All of these issues form part of ongoing work, since much more numerical data
is needed before a detailed comparison with theory is possible.

\section{Discussion}

In the above, we have classified bridge structures
in granular media as either linear or complex,
analysed their structures as obtained in our simulations,
and obtained values for some of their more important geometrical exponents.
We have also presented a theory for their orientational correlations,
which is in reasonably good agreement with our simulations,
and more importantly, makes predictions that can
be observed experimentally.
Do linear bridges really predominate at short lengths? Do they really
cross over to being domelike at large lengths,
if they survive? The answers to these and other
questions, if obtained experimentally, will not just
satisfy what is arguably mere scientific curiosity.
Our investigations above suggest that {\it long-lived bridges
are natural indicators of sustained inhomogeneities in granular systems}.
Consequently, experimental and theoretical
explorations of geometry and dynamics in stable bridge networks
will have immediate implications for issues such as
hysteresis and cooperative stability, which are some
of the most visible manifestations of complexity in granular media.

\ack

AM warmly thanks SMC-INFM
(Research and Development Center for Statistical Mechanics and Complexity,
Rome, Italy), and the Service de Physique Th\'eorique, CEA Saclay,
where parts of this work were done.
GCB acknowledges support from the Biotechnology
and Biological Sciences Research Council UK and thanks Luis Pugnaloni
for help with the figures.
Kirone Mallick is gratefully acknowledged for fruitful discussions.

\end{document}